\documentclass[aps,preprint,showpacs,nofootinbib,12pt]{revtex4}
\usepackage{graphicx}
\usepackage{epsfig}
\usepackage[dvips]{color}

\begin{document}

\title{Determining  the $\eta-\eta'$ mixing by the newly measured $BR(D(D_s)\to\eta(\eta')+\bar l+\nu_l$)}

\author{ Hong-Wei Ke$^{1}$   \footnote{khw020056@hotmail.com},
         Xue-Qian Li $^{2}$  \footnote{lixq@nankai.edu.cn} and
        Zheng-Tao Wei$^{2}$ \footnote{weizt@nankai.edu.cn}
  }

\affiliation{
  $^{1}$ School of Science, Tianjin University, Tianjin 300072, China \\
  $^{2}$ School of Physics, Nankai University, Tianjin 300071, China
  }

\begin{abstract}
\noindent The mixing of $\eta-\eta'$ or $\eta-\eta'-G$ is of a great
theoretical interest, because it concerns many aspects of the
underlying dynamics and hadronic structure of pseudoscalar mesons
and glueball. Determining the mixing parameters by fitting data is
by no means trivial. In order to extract the mixing parameters from
the available processes where hadrons are involved, theoretical
evaluation of hadronic matrix elements is necessary. Therefore
model-dependence is somehow unavoidable. In fact, it is impossible
to extract the mixing angle from a unique experiment because the
model parameters must be obtained by fitting other experiments.
Recently $BR(D\to\eta+\bar l+\nu_l)$ and $BR(D_s\to\eta(\eta')+\bar
l+\nu_l)$ have been measured, thus we are able to determine the
$\eta-\eta'$ mixing solely from the semileptonic decays of D-mesons
where contamination from the final state interactions is absent.
Thus we hope that the model-dependence of the extraction can be
somehow alleviated. Once $BR(D\to\eta'+\bar l+\nu_l)$ is measured,
we can further determine all the mixing parameters for
$\eta-\eta'-G$. As more data are accumulated, the determination will
be more accurate. In this work, we obtain the transition matrix
elements of $D_{(s)}\to \eta^{(\prime)}$ using the light-front quark
model whose feasibility and reasonability for such processes have
been tested.

\end{abstract}

\pacs{13.20.Fc, 12.39.Ki, 14.40.Cs}

\maketitle

\section{introduction}

The mixing among pseudoscalar mesons and glueballs is of great
theoretical interests and significance for understanding the
dynamics and hadronic structures. Study on the mixing has lasted for
several decades, not only because of its importance, but also the
difficulties caused by both theoretical and experimental aspects. As
is well understood, the mixing is caused by the QCD anomaly and
related to the chiral symmetry breaking
\cite{Chao:1989yp,Gershtein:1976mv}. Definitely, one would be able
to gain a better insight into the dynamics, if the mixing parameters
are more accurately determined. Many measurements on the processes
where $\eta$ and $\eta'$ are involved, have been carried out to fix
the mixing parameters. The mixing of $\eta-\eta'$ is described in
different forms, i.e. with different eigen-bases.

In the SU(3) quark model $\eta$ and $\eta'$ are the mixtures of
$\eta_0$ and $\eta_8$ which are SU(3) singlet and octet states
respectively \cite{theta1,theta2,Scora:1995ty},
\begin{eqnarray}\label{mixing1}
 \left ( \begin{array}{ccc} \eta\\  \eta' \end{array} \right )=
 \left ( \begin{array}{ccc}
  \rm \cos\theta & \rm -\sin\theta \\
  \rm \sin\theta & \rm \cos\theta \end{array} \right )
 \left ( \begin{array}{ccc} \eta_{8} \\  \eta_{0} \end{array}\right
),
 \end{eqnarray}
and the mixing angle $\theta$ was fitted in a range of $-10^\circ$
to $-23^\circ$\cite{theta2}. Bhattacharya and Rosner's recent work
\cite{theta3} suggests that the data of the decays of $D^0, \
D^{\pm}$ and $D_s$ into two light pseudoscalar mesons seem to favor
a smaller octet-singlet mixing angle $-11.7^\circ$.

In the quark content basis,  $\eta_0$ and $\eta_8$ can be further
written as mixtures of $\eta_q$ and $\eta_s$,
\begin{eqnarray}\label{mixing2}
 \left ( \begin{array}{ccc} \eta_8\\  \eta_0 \end{array} \right )=
 \left ( \begin{array}{ccc}
 \sqrt{\frac{ 1}{3}} & -\sqrt{\frac{2}{3}}\\
  \sqrt{\frac{2}{3}} & \sqrt{\frac{1}{3}} \end{array} \right )
 \left ( \begin{array}{ccc} \eta_{q} \\  \eta_{s} \end{array}\right
),
 \end{eqnarray}
where $\eta_{q}={1\over\sqrt 2}(u\bar u+d\bar d)$ and
$\eta_{s}=s\bar s$.  In Refs. \cite{kekez,phi,Fajfer:2004mv}, the
mixing of $\eta$ and $\eta'$ is written  as
\begin{eqnarray}\label{mixing11}
 \left ( \begin{array}{ccc} \eta\\  \eta' \end{array} \right )=
 \left ( \begin{array}{ccc}
  \rm \cos\phi & \rm -\sin\phi \\
  \rm \sin\phi & \rm \cos\phi \end{array} \right )
 \left ( \begin{array}{ccc} \eta_{q} \\  \eta_{s} \end{array}\right ),
 \end{eqnarray}
where $\rm
\cos\phi=\sqrt{\frac{1}{3}}\cos\theta-\sqrt{\frac{2}{3}}\sin\theta$
and $\rm
\sin\phi=\sqrt{\frac{2}{3}}\cos\theta+\sqrt{\frac{1}{3}}\sin\theta$.
We refer this mixing as scenario-I in this work. The mixing angle
$\theta=-11.7^\circ $ \cite{theta3} corresponds  to
$\phi=43.0^\circ$ in this scenario.

In the history, there have been various ways to determine the mixing
angle(s). Feldmann et al. \cite{Feldmann:1998vh} summarized the
issue and listed several possibilities to extract the mixing
angle(s) from the experimental data. By fitting the ratio of
nonleptonic decay widths
$\frac{\Gamma[J/\psi\rightarrow\eta'\rho]}{\Gamma[J/\psi\rightarrow\eta\rho]}$
the authors  of Ref. \cite{Barnett:1996hr} obtained $\phi=(39.9\pm
2.9)^\circ$, while by fitting the decay widths of
$\Gamma(\eta'\rightarrow \rho \gamma)$ and $\Gamma(\rho\rightarrow
\eta \gamma)$,  the value of $\phi$ was set as $\phi=(35.3\pm
5.5)^\circ$. In terms of
$\frac{\Gamma[a_2\rightarrow\eta'\pi]}{\Gamma[J/\psi\rightarrow\eta\pi]}$
$\phi=(43.1\pm 3.0)^\circ$ was obtained. Using the ratio of decay
widths of
$\frac{\Gamma[D_s\rightarrow\eta'e\nu]}{\Gamma[D_s\rightarrow\eta
e\nu]}$ \cite{Brandenburg:1995qq} and the form factors given in
Ref.\cite{Ball:1995zv}  $\phi$ is fixed as $\phi=(41.3\pm
5.3)^\circ$ \cite{Feldmann:1998vh}. With the cross section of
scattering processes $\pi^- p\rightarrow \eta' n$ and $\pi^-
p\rightarrow \eta n$ one had $\phi=(36.5\pm 1.4)^\circ$
\cite{Apel:1979ic} and $\phi=(39.3\pm 1.2)^\circ$
\cite{Stanton:1979pb}. The Crystal Barrel Collaboration
\cite{Amsler:1992wm} measured the ratio of the annihilation
processes $p\bar p\rightarrow \eta'$+meson($\pi^0,\eta$ or $\omega$)
and $p\bar p\rightarrow \eta$+meson($\pi^0,\eta$ or $\omega$) and
achieved $\phi=(37.4\pm 1.8)^\circ$. Based on the ratio
$\frac{\Gamma[J/\psi\rightarrow\eta'\gamma]}{\Gamma[J/\psi\rightarrow\eta\gamma]}$
the value of $\phi$ was obtained as $(39.0\pm 1.6)^\circ$
\cite{Feldmann:1998vh}. The results seem a bit dispersive, but they
are consistent with each other for the accuracy the present
experiments can reach. Feldmann et al. obtained a weighted average
of the $\phi$ value as $(39.3\pm 1.0)^\circ$.  Moreover, is not the
end of the story, since the QCD anomaly, if it causes the mixing
between $\eta$ and $\eta'$, also results in a mixing of
$\eta,\;\eta'$ with gluonium of the quantum number $0^{-+}$.

As one extends the picture to involve glueballs, a new scenario
which we refer as  the scenario-II, was suggested in Refs.
\cite{Rosner,Ball:1995zv,Benayoun:1999fv,Cheng2009,Ambrosino2007,Mathieu:2009sg}
as
\begin{eqnarray}\label{mixing12}
 \left ( \begin{array}{ccc} \eta\\  \eta' \\G\end{array} \right )=
 \left ( \begin{array}{ccc}
  \rm \cos\phi' & -\rm \sin\phi' & 0\\
  \rm \sin\phi' \cos\phi_G& \rm \cos\phi' \cos\phi_G&\sin\phi_G\\
  -\rm \sin\phi' \sin\phi_G& -\rm \cos\phi' \sin\phi_G&\cos\phi_G\end{array} \right )
 \left ( \begin{array}{ccc} \eta_{q} \\  \eta_{s} \\g\end{array}\right ),
 \end{eqnarray}
where $|g\rangle=|\rm gluonium\rangle$ is a pure gluonium state and
the physical state $G$ was identified as $\eta(1405)$
\cite{Cheng2009}. It is noted that the definition of the mixing
angles in Ref. \cite{Cheng2009} is different from that adopted in
Ref. \cite{Ambrosino2007}. In Ref. \cite{Ambrosino:2009sc} the
$\chi^2$ scheme was used to obtain $\phi_G,\phi'$ and other
parameters by fitting the data of several radiative decays such as
$\omega\rightarrow\eta\gamma$, $\rho\rightarrow\eta\gamma$
\cite{PDG08} and $\omega\rightarrow\pi^0\gamma$. The results are
${\rm sin^2} \phi_G=0.115\pm 0.036 $ and $\phi'=(40.4\pm0.6)^\circ$
if one sets $ {\rm sin^2} \phi_G$ as a free parameter.

Recently, the branching ratios of $D_s\rightarrow \eta(\eta')
e^+\nu_e$ and $D^+\rightarrow \eta e^+\nu_e$ have been measured
\cite{Yelton:2009cm,Mitchell:2008kb} and the collaboration
\cite{Mitchell:2008kb} obtains a new upper limit of
$\mathcal{BR}(D^+\rightarrow \eta' e^+\nu_e)$. Even though the
missing energy of the produced neutrino may bring up certain
uncertainties, semi-leptonic decay modes have an obvious advantage
over the radiative and non-leptonic decays. Since the contamination
from the final state interaction is absent in semi-leptonic decays,
the theoretical calculation is more reliable and the physical
quantities extracted from data, such as the mixing angles, would be
closer to reality. This is an ideal opportunity to  determine the
structure of $\eta(\eta')$, $e.g.$ extract the mixing angle(s) of
$\eta-\eta'$ ($\eta-\eta'-G$) from data. This advantage motivates us
to study  the mixing  via the semi-leptonic decays alone.

As was indicated above, in most cases, the mixing angle and the
model parameters were not simultaneously determined by fitting
solely one type of data.
Since  all the relevant processes involve
hadron transitions, one needs to evaluate the hadronic transition
matrix elements which are fully governed by the non-perturbative
QCD, in terms of phenomenological models. Therefore extraction of
the mixing is somehow model dependent.
However, if one can determine the mixing angle(s) solely from one
type of processes, the model dependence would be alleviated because
he can just employ one phenomenological model to deal with all the
concerned reactions, e.g. the model parameters and the mixing angles
are fixed altogether. Indeed, in this way, we can expect that the
model dependence which is unavoidable, could be reduced to minimum.

Our strategy is following. We will concentrate on the study of the
semi-leptonic decays of $D$ and $D_s$. At the tree-level, only the
$d\bar d$( $s\bar s$) component of $\eta,\eta'$ contributes to the
transition $D^+\rightarrow\eta(\eta')$(
$D_s\rightarrow\eta(\eta')$). The amplitude at the quark level can
be obtained in terms of the weak effective theory, so the key point
is to calculate the hadronic transition matrix elements. Concretely,
we are going to use the light front quark model (LFQM) to evaluate
the hadronic transition matrix elements. It is believed that the
LFQM is a relativistic model which has obvious advantages for
dealing with hadronic transitions where light hadrons are involved
\cite{Jaus:1999zv,Cheng:2003sm}. The light-front wave function is
manifestly Lorentz invariant and expressed in terms of the fractions
of internal momenta of the constituents which are independent of the
total hadron momentum. Applications of this approach have been
discussed in some details by the authors of Ref.
\cite{Hwang:2006cua} and their results are in good agreement with
data. Moreover, in Ref. \cite{Ke:2009ed} we explored the structure
of $f_0(980)$ via the transition $D_s\to f_0(980)$ in the LFQM.
However, in the approach, there are a few parameters to be fixed:
the $\beta$ values in the hadron wave-functions.

In our scheme, we let the mixing angle and  $\beta$ be free
parameters, when we fit the data, we obtain the mixing angle and the
$\beta$ values simultaneously. In this way, we ``extract" the mixing
angle directly from the data on the semileptonic decays of $D$ and
$D_s$. At the present $\mathcal{BR}(D^+\rightarrow \eta e^+ \nu_e)$
and $\mathcal{BR}(D_s^+\rightarrow \eta(\eta') e^+ \nu_e)$ have been
measured, but the data are not enough to determine mixing angles of
$\eta, \eta'$ and glueball. Once $\mathcal{BR}(D^+\rightarrow \eta'
e^+ \nu_e)$ is measured in the future at BES, it would be optimistic
that one will be able to further determine the mixing structure
which is extremely valuable for getting insight into the physics
picture of light hadrons and probably the glueballs. Thus so far, we
only concern the scheme I where only $\eta-\eta'$ mixing exists.


This paper is organized as follows: after the introduction, in
section II we present the form factors of $D^+\to
\eta(\eta')$($D_s\to \eta(\eta')$) which are evaluated in the LFQM,
then we calculate the corresponding decay rates and by fitting the
data we gain the mixing angle and the model parameters altogether.
Section III is devoted to our conclusion and discussions.

\section{Calculation of the widths of $D^+\rightarrow \eta(\eta') e^+ \nu_e$ and
$D_s\rightarrow \eta(\eta') e^+ \nu_e$ in LFQM}

\begin{figure}
\begin{center}
\begin{tabular}{cc}
\includegraphics[width=9cm]{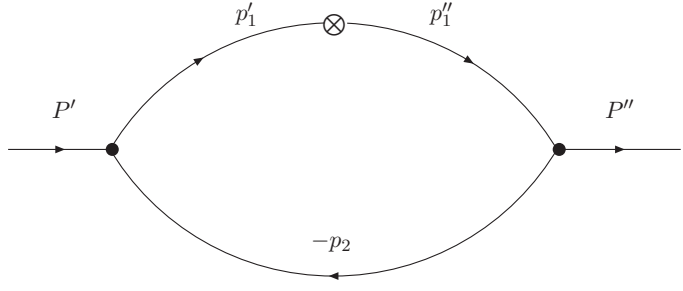}
\end{tabular}
\end{center}
\caption{ Feynman diagram for meson transition amplitude}
\label{fig:LFQM}
\end{figure}

In this section we are going to calculate the decay widths of
$D^+\rightarrow \eta(\eta') e^+ \nu_e$ and $D_s\rightarrow
\eta(\eta') e^+ \nu_e$ in terms of the LFQM. The crucial task is to
evaluate the form factors of $D^+\to \eta(\eta')$($D_s\to
\eta(\eta')$). In Ref. \cite{Wei:2009nc}, we studied $D_s\rightarrow
\eta(\eta')$ in the LFQM, thus the corresponding formulas can be
directly used in this work. The transition diagram is shown in Fig.
\ref{fig:LFQM}.

\subsection{Formulations}\label{formula}

For being self-content, we list some key formulaes given in
Refs.\cite{Jaus:1999zv,Cheng:2003sm} here. The form factors for
$P\rightarrow P$ transition are defined as
\begin{eqnarray}\label{2s1}
\langle P(P'')|A_\mu|P(P')\rangle=f_+(q^2){\mathcal{
P}}_\mu+f_-(q^2)q_\mu.
\end{eqnarray}
It is convenient to redefine them as
\begin{eqnarray}\label{2s2}
\langle P(P'')|A_\mu|P(P')\rangle=\left({
\mathcal{P}}_\mu-\frac{M'^2-M''^2}{q^2}q_\mu\right)
F_1(q^2)+\frac{M'^2-M''^2}{q^2}q_\mu F_0(q^2),
\end{eqnarray}
where $q=P'-P''$ and ${ \mathcal{P}}=P'-P''$. The relations among
the quantities are
\begin{eqnarray}
F_1(q^2)=f_+(q^2),\,\, F_0(q^2)=f_+(q^2)+\frac{q^2}{q\cdot{
\mathcal{P}}}f_-(q^2).
\end{eqnarray}

Functions $f_{\pm}(q^2)$ can be calculated in the LFQM and their
explicit expressions were presented as \cite{Cheng:2003sm},
\begin{eqnarray}\label{2s4}
f_+(q^2)=&&\frac{N_c}{16\pi^3}\int dx_2d^2p'_\perp
 \frac{h'_ph''_p}{x_2\hat{N}'_1\hat{N}''_1}
 \left[-x_1(M'^2_0+M''^2_0)-x_2q^2+x_2(m_1'-m_1''^2)\right.\nonumber\\&&
 \left.+x_1(m'_1-m_2)^2+x_1(-m''_1+m_2)^2\right],\nonumber\\
f_-(q^2)=&&\frac{N_c}{16\pi^3}\int dx_2d^2p'_\perp
\frac{2h'_ph''_p}{x_2\hat{N}'_1\hat{N}''_1}
 \left\{x_1x_2M'^2+p'_\perp+m_1'm_2+(-m_1''+m_2)(x_2m_1'+x_1m_2)\right.\nonumber\\&&
 -2\frac{q\cdot \mathcal{P}}{q^2}\left(p'^2_\perp+2\frac{(p'_\perp\cdot
 q_\perp)^2}{q^2}\right) -2\frac{(p'_\perp\cdot
 q_\perp)^2}{q^2}+\frac{(p'_\perp\cdot
 q_\perp)}{q^2}\left[M''^2-x_2(q^2+q\cdot \mathcal{P})\right.\nonumber\\
 &&-(x_2-x_1)M'^2 \left.\left.
 +2x_1M'^2_0-2(m'_1-m_2)(m'_1+m''_1)\right]\right\}.
\end{eqnarray}
where $m'_1,\; m''_1$ and $m_2$ are the corresponding quark masses,
$M'$ and $M''$ are the masses of the initial and final mesons
respectively. All other notations can be found in the Appendix.

The form factors are parameterized in a three-parameter form as
 \begin{eqnarray}\label{s14}
 F(q^2)=\frac{F(0)}{1-a\left(\frac{q^2}{M^2}\right)
  +b\left(\frac{q^2}{M^2}\right)^2}.
 \end{eqnarray}
where $F(q^2)$ represents the form factors $F_1,~ F_0$, and $F(0)$
is the form factor at $q^2=0$; $M$ is the mass of the initial meson.
The three parameters $F(0)$ and $a,~b$ are fixed by performing a
three-parameter fit to the form factors  which are calculated in the
space-like region and then extended to the physical time-like
region.

For semileptonic decay of a pseudoscalar meson ($D$ or $D_s$) into a
pseudoscalar meson, i.e. $P(P')\to P(P'')l\nu$, the differential
width is \cite{Zweber:2007zz}
 \begin{eqnarray}
 \frac{d\Gamma}{dq^2}(P\to Pl\nu)=\frac{G_F^2|V_{CKM}|^2~p^3}{24\pi^3}|F_1(q^2)|^2,
 \end{eqnarray}
where $q=P'-P''$ is the momentum transfer and $q^2$ is the invariant
mass of the lepton-neutrino pair; $p$ is the final meson momentum in
the $D$ or $D_s$ rest frame and
 \begin{eqnarray}
 p=|\vec{P''}|=\frac{\sqrt{\left(M^2-(M_f-\sqrt{q^2})^2\right)
  \left(M^2-(M_f+\sqrt{q^2})^2\right)}}{2M}.
\end{eqnarray}
$M_f$ denotes the mass of the produced meson. It is noted that the
differential  width is governed by only one form factor $F_1(q^2)$
because we neglect the light lepton masses.

\subsection{ Calculating the decay rates of
 $D^+(D_s)\to \eta(\eta')e^+\nu_e$ and determining the mixing angle}

Now, let us extract the mixing angle in scenario-I by means of the
experimental data of $\mathcal{BR}(D^+\rightarrow \eta
e^+\nu_e)=(13.3\pm2.0\pm0.6)\times 10^{-4}$,
$\mathcal{BR}(D_s\rightarrow\eta e^+
\nu_e)=(2.48\pm0.29\pm0.13)\times10^{-2}$ and
$\mathcal{BR}(D_s\rightarrow\eta' e^+
\nu_e)=(0.91\pm0.33\pm0.05)\times10^{-2}$.

First we need to calculate the form factors
$F_1^{D^+(D_s)\eta}(q^2)$ and $F_1^{D^+(D_s)\eta'}(q^2)$ using the
formulas given in section \ref{formula}. Since $d$ quark in $D^+$
appears in the final meson, only the $d\bar{d}$ component of
$\eta$($\eta'$) contributes to the transition of $D^+ \to
\eta(\eta')\bar l\nu_l$ $i.e.$ the mixing angle is included in the
form factors. The similar situation for the transition $D_s \to
\eta(\eta')\bar l\nu_l$ was discussed in \cite{Wei:2009nc} but we
let the mixing angle be free.


\begin{figure}
\begin{center}
\begin{tabular}{ccc}
\includegraphics[scale=0.7]{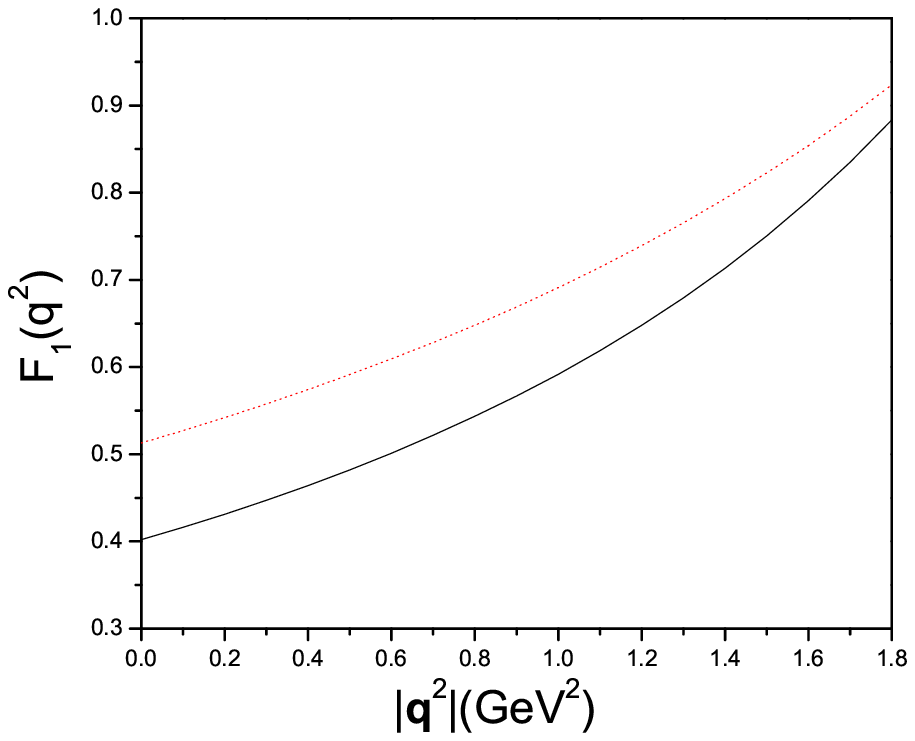}
\includegraphics[scale=0.7]{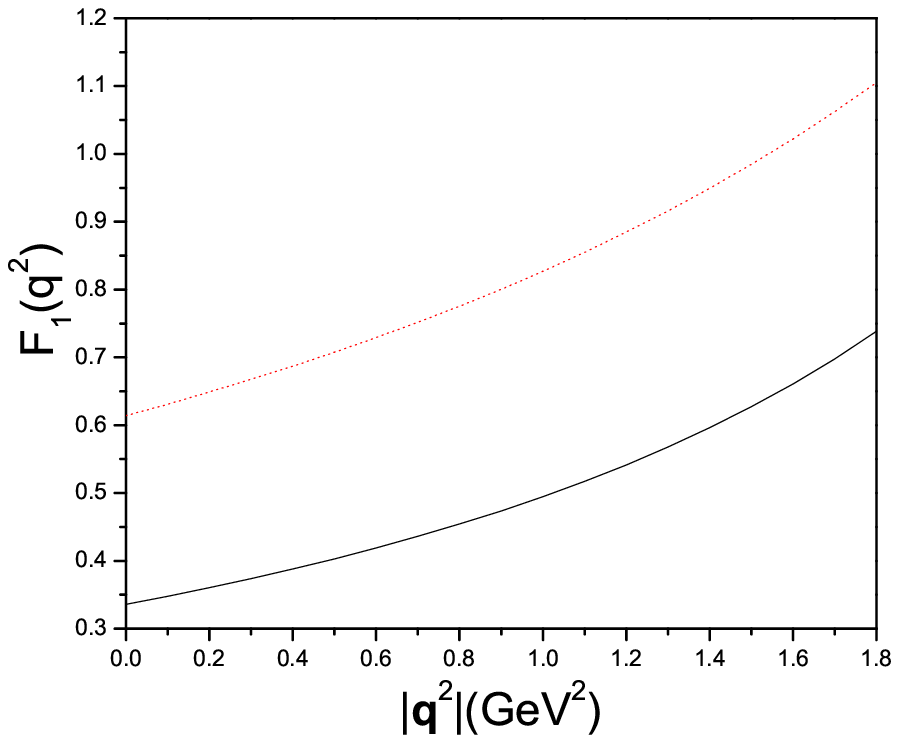}
\end{tabular}
\end{center}

\caption{The dependence of the form factor $F_1(q^2)$ on $q^2$ at
$\phi=39.9^\circ$. (a) $D^+\rightarrow \eta$ (the solid line) and
$D_s\rightarrow \eta$ (the dotted line); (b) $D^+\rightarrow
\eta'$ (the solid line) and $D_s\rightarrow \eta'$ (the dotted
line) } \label{fig:fbeta}
\end{figure}

In this work, the input parameters for quark masses are directly
taken from \cite{Cheng:2003sm} as $m_u=0.26$ GeV, $m_s=0.37$ GeV,
$m_c=1.4$ GeV. From the experimental results for decay constants
$f_{D}^{\rm exp}=0.221~{\rm MeV}, ~f_{D_s}^{\rm exp}=0.27~ {\rm
MeV}$, the parameters for $\beta_{D_s}$ and $\beta_D$ are fixed to
be $\beta_{D_s}=0.592$ GeV, $\beta_{D}=0.499$ GeV
\cite{Wei:2009nc}.

Totally, there are five free parameters $\beta^q_\eta$,
$\beta^q_{\eta'}$, $\beta^s_\eta$, $\beta^s_{\eta'}$ and $\phi$ to
be fixed. It seems that we do not have enough equations to
determinate all these parameters. However, the $\beta_P^{q\;{\rm
or}\; s}$ is a parameter in the Gaussian wave function whose value
only depends on the quark flavor (q or s), but not on the characters
of the hadron, therefore two relations
$\beta^q_\eta=\beta^q_{\eta'}$ and $\beta^s_\eta=\beta^s_{\eta'}$
hold. The relations are also deduced in the Appendix B where we do
not make any special assumption on them. By these relations we
reduce the number of unknowns to three. From the three decay modes,
we obtain the values of $\phi$ and $\beta_{\eta(\eta')}^q, ~
\beta_{\eta(\eta')}^s$. The mixing angle is fit to be
$\phi=(39.9\pm2.6)^\circ$ where the errors are from the experiment.
The parameters $\beta$ are $\beta_{\eta(\eta')}^q=0.398$ GeV,
$\beta_{\eta(\eta')}^s=0.453$ GeV.

Now, let us discuss the theoretical uncertainties in our used light
front quark model. Here, we only address the errors coming from the
phenomenological parameters. The numerical results are not sensitive
to the variance of $\beta_{\eta(\eta')}^q$ and
$\beta_{\eta(\eta')}^s$. In \cite{Wei:2009nc}, it is found that the
variations of decay constants of $f_D$ and $f_{D_s}$ with parameters
$\beta$ is nearly a linear relation. One can ascribe the
uncertainties from $\beta_D, ~\beta_{D_s}$ to experimental errors of
$f_D, ~f_{D_s}$. For the quark masses, we made a variation of their
values by 10\% and explore the uncertainties of predictions. We
found that the numerical numbers are not sensitive to $m_d,~ m_s$.
In particular, with variations of $m_d,~ m_s$, the errors of $\phi$
is only $0.7^\circ$. The uncertainties from charm quark mass $m_c$
is in a controllable region, $\Delta\phi=2.1^\circ$. Combing them,
we have an error $2.3^\circ$ in determination of $\phi$, this is the
main error of our approach. After considering the experimental
error, the mixing angle is determined to be
$\phi=\left(39.9\pm2.6({\rm exp})\pm2.3({\rm the} )\right)^\circ$.

SU(3) breaking is a substantial effect in $\eta-\eta'$ mixing. In
our approach, we have included their effects in differences of quark
masses $m_d$ and $m_s$; $\beta$ parameter differences in
$\beta_\eta^q$ and $\beta_\eta^s$. Using these values, we plot the
dependence of the form factor $F_1(q^2)$ on $q^2$ in the physical
region of $q^2\geq 0$ in Fig. 2 and we estimate
$\mathcal{BR}(D^+\rightarrow \eta' e^+\nu_e)=(2.12\pm0.23({\rm
exp})\pm 0.20({\rm the}))\times 10^{-4}$ which is lower than the
experimental upper bound $\mathcal{BR}(D^+\rightarrow \eta'
e^+\nu_e)\leq 3.5\times 10^{-4}$ \cite{Mitchell:2008kb}.

Using the formula of Eq. (\ref{bpp1}) in  Appendix and the above
parameters we calculate the central value of the decay constants
$f^q_\eta=80$ MeV, $f^s_\eta=113$ MeV, $f^q_{\eta'}=67$ MeV and
$f^s_{\eta'}=145$ MeV. These values are obtained by fit to the
experiment. They are very close to the results $f^q_\eta=78$ MeV,
$f^s_\eta=113$ MeV, $f^q_{\eta'}=64$ MeV and $f^s_{\eta'}=141$ MeV
given in Ref.\cite{Ali:1998eb} where two mixing angles
$\theta_0=-9.1^\circ$ and $\theta_8=-22.2^\circ$ are used for the
mixing of $\eta$ and $\eta'$. It is noted that our result extracted
solely from the data on the semi-leptonic decays of $D$ and $D_s$ is
consistent with the previously determined weighted value
$39.3^\circ$ \cite{Feldmann:1998vh} and $41.4^\circ$
\cite{Ambrosino:2009sc} within a reasonable error tolerance. In
Refs. \cite{Ball:1995zv,Feldmann:1998vh} the authors also used the
data $\Gamma(D_s\rightarrow \eta' e\nu)/\Gamma(D_s\rightarrow\eta
e\nu)$ to determine the mixing angle. In this work, we employ an
alternative model to evaluate the hadronic transition matrix
elements. Moreover, nowadays there are more data available, which
enable us to make more accurate estimation on the $\eta-\eta'$
mixing.

In principle we will be able to fix the mixing angles $\phi'$ and
$\phi_G$ in Eq.(\ref{mixing12}) simultaneously as long as  a
sufficiently large database on the semileptonic decay modes is
available.

\section{Conclusion}

Because of absence of the final state interactions, the semileptonic
decays have obvious advantages for determining the properties of the
produced light hadrons, such as the structure of $f_0(980)$,
$\eta-\eta'$ mixing and even a mixing of pseudoscalar mesons with
glueball, over other modes from the theoretical aspect. Indeed,
moreover, for the semileptonic decays, one only needs to calculate
the hadronic transition elements where only one final hadron is
involved. Instead, besides the complication caused by the final
state interaction, the evaluation of hadronic matrix elements of the
non-leptonic decays is much more difficult because there are two
hadrons in the final state. To do the job, the factorization scheme
may be invoked and more ambiguities are raised by the scheme.

In this work we extract the mixing angle of $\eta$ and $\eta'$ by
the semileptonic decays $D\rightarrow \eta e^+\nu_e$ and
$D_s\rightarrow \eta(\eta') e^+\nu_e$. At first we calculate the
form factors of $D\rightarrow \eta $ and $D_s\rightarrow \eta(\eta')
$ in the LFQM where the mixing angle  of $\eta$ and $\eta'$ and the
parameters $\beta$s were free.  Then we  compute the rates of the
semileptonic decays $D\rightarrow \eta e^+\nu_e$ and $D_s\rightarrow
\eta(\eta') e^+\nu_e$. Equating our theoretical derivations with the
data we obtain $\phi=(39.9\pm2.6({\rm exp})\pm2.3({\rm
the}))^\circ$, $\beta_{\eta(\eta')}^q=0.398$ GeV and
$\beta_{\eta(\eta')}^s=0.453$ GeV.  We estimate
$\mathcal{BR}(D^+\rightarrow \eta' e^+\nu_e)=(2.12\pm0.23({\rm
exp})\pm0.20({\rm the}))\times 10^{-4}$.

As indicated above, there are three free parameters including the
$\eta-\eta'$ mixing in our model-dependent calculations, and we
fix them simultaneously by the data of $BR(D\rightarrow \eta
e^+\nu_e)$ and $BR(D_s\rightarrow \eta(\eta') e^+\nu_e)$ in our
scheme I. If in the future $\mathcal{BR}(D^+\rightarrow \eta'
e^+\nu_e)$  can be accurately measured (not just an upper bound),
we will be able to test our theoretical prediction of
$(2.12\pm0.23(\rm exp)\pm0.20(\rm the))\times 10^{-4}$ which is
made in our scheme I. If the data are consistent with our
estimation, it implies that our scheme I is valid, e.g the mixing
between $\eta-\eta'$ with gluonium states is small. Instead, if
there is an observable discrepancy between our theoretical
prediction on $\mathcal{BR}(D^+\rightarrow \eta' e^+\nu_e)$ and
the data, one needs to further consider the scheme II where the
mixing between $\eta-\eta'$ is accounted, and by the data, we can
determine the mixing angles $\phi'$ and $\phi_G$ in eq.(4). We
wish our experimental colleagues to carry out measurements on
$D^+\rightarrow \eta' e^+\nu_e$ and further improve the accuracy
of the measurements of semileptonic decays.

Of course, the branching ratios of semileptonic decay modes are
smaller than that of non-leptonic modes, and due to the existence
of a neutrino (antineutrino), the missing energy would make the
event re-construction more difficult. However, the obvious
advantage of the semileptonic decays for determining the
properties of the light hadrons, would be the reason to urge our
experimental colleagues to conduct more precise measurements. It
is worthwhile.

\section*{Acknowledgments}

This work is supported by the National Natural Science Foundation
of China (NNSFC) under the contract No. 10775073, No. 10705015 and
the special grant for the PH.D program of the Chinese Education
Ministry. One of us (Ke) is also partly supported by the special
grant for new faculty from Tianjin University.

\appendix

\section{Notations}

Here we list some variables appearing in the context.  The incoming
(outgoing) meson in Fig. \ref{fig:LFQM} has the momentum
${P'}^({''}^)={p_1'}^({''}^)+p_2$ where ${p_1'}^({''}^)$ and $p_2$
are the momenta of the off-shell quark and antiquark and
\begin{eqnarray}\label{app1}
&& p_1'^+=x_1P'^+, \qquad ~~~~~~p_2^+=x_2P'^+, \nonumber\\
&& p'_{1\perp}=x_1P'_{\perp}+p'_\perp, \qquad
 p_{2\perp}=x_2P'_{\perp}-p'_\perp,
 \end{eqnarray}
with  $x_i$ and $p'_\perp$ are internal variables and $x_1+x_2=1$.

The variables $M_0'$, $\tilde {M_0'}$, $h_p'$ and $\hat{N_1'}$ are
defined as
\begin{eqnarray}\label{app2}
&&M_0'^2=\frac{p'^2_\perp+m'^2_1}{x_1}+\frac{p'^2_\perp+m^2_2}{x_2},\nonumber\\&&
\tilde {M_0'}=\sqrt{M_0'^2-(m_1'-m_2)^2}.
 \end{eqnarray}

\begin{eqnarray}\label{app3}
h_p'=(M'^2-M'^2_0)\sqrt{\frac{x_1x_2}{N_c}}\frac{1}{\sqrt{2}\tilde{M_0'}}\varphi',
 \end{eqnarray}
where
\begin{eqnarray}\label{app4}
\varphi'=4(\frac{\pi}{\beta'^2})^{3/4}\sqrt{\frac{dp_z'}{dx_2}}{\rm
 exp}(-\frac{p'^2_z+p'^2_\perp}{2\beta'^2}),
 \end{eqnarray}
with $p_z'=\frac{x_2M_0'}{2}-\frac{m_2^2+p'^2_\perp}{2x_2M_0'}$.
\begin{eqnarray}\label{app5}
\hat{N_1'}=x_1(M'^2-M'^2_0).
 \end{eqnarray}

\section{the relations between $\beta^{q}_{\eta(\eta')}$ and $\beta^{s}_{\eta(\eta')}$}
For a pseudoscalar meson the decay constant can be evaluated
\begin{eqnarray}\label{bpp1}
f_P=\frac{\sqrt{N_c}}{16\pi^3}\int
dx_2d^2p'_\perp\frac{\varphi'}{\sqrt{2x_1x_2}\tilde
{M_0'}}4(m_1'x_2+m_2x_1),
 \end{eqnarray}
so the parameters $\beta^{q}_{\eta^{(')}}$, $\beta^{s}_{\eta^{(')}}$
can be determined by the decay constants of $f^{q}_{\eta(\eta')}$
and $f^{s}_{\eta(\eta')}$ which are defined in Ref.
\cite{Feldmann:1998vh}
\begin{eqnarray}
&& f_\eta^{q} = f_{q} \, \cos\phi \ , \qquad
    f_\eta^{s} = - f_{s} \, \sin\phi \ ,\nopagebreak \cr\nopagebreak
&&  f_{\eta'}^{q} = f_{q} \, \sin\phi \ , \qquad
   f_{\eta'}^{s} =  \ \ f_{s} \, \cos\phi \ .
\label{bpp2}
\end{eqnarray}
where $q$ represents $u$ or $d$ quark.

Since  $\eta=\eta_q\cos\phi -\eta_s\sin\phi$ and
$\eta'=\eta_q\sin\phi +\eta_s\cos\phi$, we get the relations
$\beta^{q(s)}_{\eta}=\beta^{q(s)}_{\eta'}$

\end{document}